\documentclass[12pt,preprint]{aastex}

\slugcomment{Accepted for Publication in the Astronomical Journal}

\shorttitle{Grain Growth in the Dark Cloud L1251}
\shortauthors{Kandori et al.}

\begin{document}

\title{Grain Growth in the Dark Cloud L1251}

\author{Ryo Kandori\altaffilmark{}} 
\affil{Department of Astronomical Science, Graduate University for Advanced Studies (Sokendai), Osawa, Mitaka, Tokyo 181-8588, Japan}

\email{kandori@alma.mtk.nao.ac.jp}

\author{Kazuhito Dobashi\altaffilmark{}, Hayato Uehara\altaffilmark{}, and Fumio Sato\altaffilmark{}} 
\affil{Department of Astronomy and Earth Sciences, Tokyo Gakugei University, Koganei, Tokyo 184-8501, Japan}

\email{dobashi@u-gakugei.ac.jp, hayato.uehara@mac.com, satofm@u-gakugei.ac.jp}

\and

\author{Kenshi Yanagisawa\altaffilmark{}}
\affil{Okayama Astrophysical Observatory, National Astronomical Observatory of Japan, Kamogata, Okayama 719-0232, Japan}

\email{yanagi@oao.nao.ac.jp}

\begin{abstract}
We have performed optical imaging observations of the dark cloud L1251 at multiple wavelengths, $B$, $V$, $R$, and $I$, using the 105 cm Schmidt telescope at the Kiso Observatory, Japan. The cloud has a cometary shape with a dense ``head" showing star formation activity and a relatively diffuse ``tail" without any signs of star formation. We derived extinction maps of $A_{B}$ and $A_{V}$ with a star count method, and also revealed the color excess ($E_{B-V}$, $E_{V-R}$, and  $E_{V-I}$) distributions. On the basis of the color excess measurements we derived the distribution of the ratio of total to selective extinction $R_{V}$ over the cloud using an empirical relation between $R_{V}$ and $A_{\lambda}$/$A_{V}$ reported by Cardelli et al. In the tail of the cloud, $R_{V}$  has a uniform value of $\sim$3.2, close to that often found in the diffuse interstellar medium ($\sim$3.1), while higher values of $R_{V}$=4$-$6 are found in the dense head.  Since $R_{V}$  is closely related to the size of dust grains, the high $R_{V}$-values are most likely to represent the growth of dust grains in the dense star-forming head of the cloud.
\end{abstract}


\section{Introduction} 
The ratio of total to selective extinction, $R_{V}$=$A_{V}$/$E_{B-V}$, is a measure of wavelength dependence of the interstellar extinction, representing the shape of the extinction curve. $R_{V}$ is an important parameter to characterize dust properties at optical wavelengths because it is closely related to the size and composition of dust grains (e.g., Mathis, Rumpl, \& Nordsieck 1977; Hong \& Greenberg 1978; Kim, Martin, \& Hendry 1994). It is well established that $R_{V}$ has a uniform value of $\sim$3.1 in the diffuse interstellar medium (e.g., Whittet 1992), but it tends to be much higher in dense molecular clouds ($R_{V}$=4$-$6; e.g., Vrba, Coyne, \& Tapia 1993; Strafella et al. 2001). Many theoretical predictions attribute the higher $R_{V}$ in the dense cloud interior to a significant change of dust properties, particularly in size (e.g., Mathis 1990;  Ossenkopf \& Henning 1994).  Since the timescale of grain growth due to coagulation of smaller grains and/or accretion of molecules in the gas phase onto the larger grain surface depends greatly on density (e.g., Tielens 1989), the higher $R_{V}$-values observed in dense molecular clouds are most likely to represent the growth of dust grains. \par
There have been a number of observational reports on $R_{V}$ measurements to date (e.g., Larson et al. 2000; Whittet et al. 2001). However, they are mostly based on the spectrophotometry of a limited number of stars lying in the background of the clouds, and the $R_{V}$-values have been measured only toward the selected stars. A large-scale distribution of $R_{V}$ over an entire cloud, as well as its local variation inside, are therefore still poorly known.  \par
In order to investigate how much $R_{V}$ can vary in a single cloud, we have carried out imaging observations of L1251 at multiple wavelengths, $B$, $V$, $R$, and $I$, using the 105 cm Schmidt telescope at the Kiso Observatory, Japan. This cloud is located in the Cepheus Flare ($\alpha$$\sim$$22^{h}35^{m}$, $\delta$$\sim$$75^{\circ}00'$; J2000.0) and has an elongated shape. Since L1251 is an isolated dark cloud located at a relatively high Galactic latitude ($\sim$$15^{\circ}$), it is expected that the cloud is free from contamination due to unrelated clouds in the same line of sight. Kun \& Prusti (1993) estimated its distance to be 300$\pm$50 pc with a color excess measurement and spectral classification of stars. They also found 12 H${}_{\alpha}$ emission-line stars at the eastern part of the cloud, five of which are detected as $IRAS$ point sources. Global molecular distribution of the cloud was first revealed by Sato \& Fukui (1989) and Sato et al. (1994) through ${}^{13}$CO and C${}^{18}$O ($J$=1$-$0) observations carried out with the 4 m radio telescope at the Nagoya University (HPBW=2.\hspace{-2pt}$'$7). Their observations clearly revealed the cometary ``head-tail" morphology of L1251. They also found that two of the $IRAS$ point sources located in the eastern ``head" part of the cloud are associated with molecular outflows, while no sign of ongoing star formation was found in the western ``tail." These characteristics of L1251 make this cloud a suitable site to investigate the possible difference in the dust properties (e.g., $R_{V}$) between star-forming and non-star-forming regions in a single cloud. \par
In this paper, we report results of the $BVRI$ observations toward L1251. The observational procedures are described in $\S$ 2.  The distribution of extinction ($A_{B}$ and $A_{V}$) and color excess at each band ($E_{B-V}$, $E_{V-R}$, $E_{V-I}$) are derived from the obtained data, which reveal the global dust distribution of the cloud. These are shown in $\S\S$ 3 and 4. We also derive $R_{V}$  distribution over the cloud on the basis of color excess measurements and an empirical relation between $R_{V}$  and $A_{\lambda}$/$A_{V}$ (Cardelli, Clayton, \& Mathis 1989). A significant difference has been found in $R_{V}$ distribution between the dense head and the relatively diffuse tail of the cloud, suggesting a larger dust size in the dense head than in the tail. We introduce our procedure to create $R_{V}$ map and compare the obtained $R_{V}$ distribution with the other data in $\S$ 5. Our conclusions are summarized in $\S$ 6. 

\section{Observations and Data Reduction}
We carried out imaging observations of L1251 from 2000 November 27 to December 2 using the 105 cm Schmidt telescope equipped with the 2kCCD camera at the Kiso Observatory. The camera has a backside-illuminated 2048 $\times$ 2048 pixel chip covering a wide field of view of $\sim$$50'$$\times$$50'$ with a scale of 1.\hspace{-2pt}$''$5 pixel${}^{-1}$ at the focal plane of the telescope.  Using the camera, we mapped L1251 with the $B$, $V$, $R$, and $I$ filters at a $40'$ grid spacing along the equatorial coordinates, so that adjacent frames could have an overlap of $\sim$10$'$ on the sky. We performed integrations of 200, 100, 100, and 50 s for the images obtained with the $B$, $V$, $R$, and $I$ filters, respectively. The typical seeing was 3.\hspace{-2pt}$''$0$-$4.\hspace{-2pt}$''$5 during the observations. In total, we obtained 10 frames per filter, covering the entire extent of the cloud.  \par
To calibrate the images, we obtained dome flat frames in the beginning of the observations and frequently obtained bias frames during the observations. In addition, we observed some sets of standard stars listed by Landolt (1992) to correct the atmospheric extinction and calibration.  \par
We processed the calibrated images in a standard way using the IRAF packages.\footnote{IRAF is distributed by the National Astronomy Observatory, which is operated by the Association of Universities for Research in Astronomy, Inc., under coorperative agreement with the National Science Foundation.} After replacing values at the defected pixels in the detector (i.e., dead pixels) by those interpolated using surrounding pixels, we applied the stellar photometry package DAOPHOT (Stetson 1987) to the images to detect stars and then to make photometric measurements.  To improve the photometry we further measured the instrumental magnitudes of the detected stars by fitting a point-spread function (PSF) and then calibrated them for the atmospheric extinction. Among all of the detected stars we selected the ones having an intensity greater than $4.0$ $\sigma$ noise level of the sky background. Some false detection, however, could not be completely ruled out, which we later removed by eye inspection.  \par
For calculating the plate solution, we compared the pixel coordinates of a number of the detected stars with the celestial coordinates of their counterparts in the Digitized Sky Survey (DSS), which resulted in a positional error of  less than 1$''$ (rms). All of the stars detected in our observations were summarized separately with respect to the filters we used ($BVRI$), which we will adopt to generate an extinction map using a star count method. The limiting magnitudes are 20.0, 19.7, 19.0, and 18.3 mag for the $B$, $V$, $R$, and $I$ bands, respectively. Among all of the identified stars $\sim$33,000 were detected in all of the four bands. These stars are useful for probing the distribution of the color excess in the cloud.  We converted the instrumental magnitudes of these stars into the magnitudes in the standard Johnson-Cousins photometric system (Johnson \& Morgan 1953; Cousins 1976), and we also measured their colors ($B$$-$$V$, $V$$-$$R$, and $V$$-$$I$). For the color excess measurements, we constructed a $BVRI$-matched star list for $\sim$30,600 stars whose photometric errors in the color measurements, $\sigma_{V-\lambda}$, are less than 0.15 mag.
One should note that we excluded all of the known young stellar objects (Sato \& Fukui 1989; Sato et al.1994; Kun \& Prusti 1993) cross-identified in our photometries in order to avoid errors in the following analyses since they are unrelated to the global extinction in L1251.

\section{Distribution of Extinction}
On the basis of the stars detected in the $B$ and $V$ band, we derived extinction (i.e., $A_{B}$ and $A_{V}$) maps of L1251 using a star count method. Given a cumulative stellar density in the extinction-free reference field $N_{{\rm ref}}$ as a function of the $\lambda$-band magnitude $m_{\lambda}$ (i.e., the Wolf diagram; Wolf 1923), a logarithmic cumulative stellar density log $N$ measured at the threshold magnitude $m_{\lambda,0}$ can be converted into the extinction $A_{\lambda}$ as
\begin{eqnarray}
{A}_{\lambda}\left({l, b, {m}_{\lambda,0}}\right) = {m}_{\lambda,0} - {f}^{- 1} \left[ {\log N\left({l, b, {m}_{\lambda,0}}\right)} \right]\hspace{2mm},
\end{eqnarray}
where $l$ and $b$ are the Galactic coordinates, and  $f^{-1}$ is the inverse function of $f$ defined as log $N_{{\rm ref}}$ = $f(m_{\lambda})$. The function $f$ is often assumed to be linear as $f=a+b{m}_{\lambda}$, where $b=\left[{d\log N_{{\rm ref}} / dm_{\lambda} }\right]_{m_{\lambda}=m_{\lambda,0}}$ (e.g., Dickman 1978). \par
To derive $N(l, b, m_{\lambda,0})$ in equation (1) from the obtained data, we first set circular cells with a 4$'$ diameter spaced by 2$'$ along the Galactic coordinates, and we counted the number of stars in the cells, setting the threshold magnitude to be $m_{B,0}$ = 19.5 mag and $m_{V,0}$ = 19.2 mag, which is well above the detection limit in our observations. We then smoothed the stellar density map to a 6$'$ resolution with Gaussian filtering for a better signal-to-noise ratio to produce the final stellar density map $N(l, b , m_{\lambda,0})$, which we adopted to derive the $A_{B}$ and $A_{V}$ distribution. After smoothing typical stellar density in the reference field and in the most opaque region was $\sim$20 cell${}^{-1}$ and $\sim$1 cell${}^{-1}$ for B band and $\sim$30 cell${}^{-1}$ and $\sim$2 cell${}^{-1}$ for V band, respectively. \par
We composed the stellar luminosity function $f_{{\rm ref}}(m_{\lambda})$ in the reference field, i.e., the area expected to be free from extinction in the observed region (see Fig. 1). We found that the function is actually not linear, but has a curved shape. We also found that stellar densities at the threshold magnitude $m_{\lambda,0}$ systematically decrease along with the Galactic latitudes $b$. To deal with this problem, we fitted the average luminosity function $f_{{\rm ref}}(m_{\lambda})$ by a 4th-order polynomial and scaled it to the background stellar density $N_{{\rm back}}$ at $m_{\lambda}$$=$$m_{\lambda,0}$ as
\begin{eqnarray}
f\left({l, b, {m}_{\lambda}}\right) = {\frac{\log {N}_{\rm back}\left({l, b, {m}_{\lambda,0}}\right)}{{f}_{\rm ref}\left({{m}_{\lambda,0}}\right)}}\sum\limits_{ n = 1}^{ 4} { c}_{ n}{ m}_{\lambda}^{ n}\hspace{2mm},
\end{eqnarray}
where $c_{n}$ represents the best-fitting coefficients. The value $f_{{\rm ref}}(m_{\lambda,0})$ corresponds to the mean stellar density at $m_{\lambda,0}$ in the reference field. We determined $N_{{\rm back}}(l, b, {m}_{\lambda,0})$ by assuming an exponential dependence on $b$ as
\begin{eqnarray}
{N}_{{\rm back}}\left({l, b, {m}_{\lambda,0}}\right) = {N}_{0}\exp \left({- \alpha \left|{b}\right|}\right)\hspace{2mm},
\end{eqnarray}
where $N_{0}$ and $\alpha$ are constants determined using values of $N(l, b, m_{\lambda,0})$ in the reference field. We obtained the coefficients (${N}_{0}$, $\alpha$) to be (61.5, 0.07) for $B$ band and (106.2, 0.09) for $V$ band. \par
We numerically calculated $f^{-1}$, the inverse function of $f$ in equation (2), and used it in equation (1) to derive $A_{B}$ and $A_{V}$ after substituting the actually measured stellar density $N(l, b , m_{\lambda,0})$ into $f^{-1}$. The resulting $A_{V}$ map is shown in Figure 1. The distribution of $A_{B}$ (not shown) is very similar to that of $A_{V}$. Total uncertainty in $A_{\lambda}$ varies from region to region, depending on the number of stars falling in the cells. Typical uncertainties in the reference field ($A_{\lambda}$ $\sim$ 0 mag) and in the most opaque region are estimated to be $\sim$0.5 and $\sim$0.8 mag for the $B$ band, and $\sim$0.35 and $\sim$0.7 mag for the $V$ band, respectively. \par
The $A_{V}$ map reveals an interesting dust distribution in L1251 with a cometary head-tail morphology, which is similar to the molecular distribution traced in ${}^{13}$CO (Sato et al. 1994).

\section{Distribution of Color Excess}
We derived the color excess distribution in L1251 using the near-infrared color excess (NICE) method originally developed by Lada et al. (1994). The method compares the color index of stars in the reference field with those in the dark clouds. If we assume that the population of the stars across the observed field is invariable, the mean stellar color in the reference field can be used to approximate the mean intrinsic color of stars (i.e., $<$$V$$-$$\lambda$$>$${}_{{\rm int}}$$=$$<$$V$$-$$\lambda$$>$${}_{{\rm ref}}$), and the color excess can be derived by subtracting $<$$V$$-$$\lambda$$>$${}_{{\rm ref}}$ from the observed color of stars. To obtain the color excess maps, we adopted an ``adaptive grid" technique (Cambr\'{e}sy 1998, 1999) to the $\sim$30,600 stars detected in all of the four bands ($B$, $V$, $R$, and $I$). This technique consists in fixing the number of counted stars $N$ per variable cell to achieve a uniform noise level in the resulting map. To estimate the mean color excess $<$$E$${}_{V-\lambda}$$>$ ($\lambda$$=$$B$, $R$, or $I$) in circular cells spaced by 2$'$ along the Galactic coordinates with undefined sizes, we calculated the mean color of the closest $N$ stars to the center positions of the cells by setting $N$ to be 10 and derived their shifts from the mean color in the reference field $<$$V$$-$$\lambda$$>$${}_{{\rm ref}}$ as
\begin{eqnarray}
\left\langle{{E}_{V -\ \lambda }}\right\rangle\left({l,b}\right) = \left[ {\sum\limits_{i = 1}^{N} {\frac{{\left({V -\ \lambda }\right)}_{i}}{N}}} \right]\left({l,b}\right) - {\left\langle{V -\ \lambda }\right\rangle}_{{\rm ref}}\hspace{2mm},
\end{eqnarray}
where  ($V$$-$$\lambda$)${}_{i}$ is the color index of the $i$th star in a cell. The distance to the $N$th star from the cell center gives the radius of the cell, corresponding to one half of the angular resolution. $<$$V$$-$$\lambda$$>$${}_{{\rm ref}}$ is the mean color of the $\sim$2000 cells in the reference field, and we determined $<$$B$$-$$V$$>$${}_{{\rm ref}}$, $<$$V$$-$$R$$>$${}_{{\rm ref}}$, and $<$$V$$-$$I$$>$${}_{{\rm ref}}$ to be 1.02, 0.55, and 1.23 mag, respectively.  \par
In the reference field, we found that $<$$V$$-$$\lambda$$>$ slightly but systematically decreases with increasing Galactic latitudes $b$. This may be caused by the presence of diffuse dust layer of the Galaxy and/or gradual change in the stellar population toward the Galactic plane. We calibrated the effect by substituting $<$$V$$-$$\lambda$$>$${}_{{\rm ref}}$ in equation (4) for the following equation assuming an exponential dependence on $b$ as
\begin{eqnarray}
{\left\langle{V -\ \lambda }\right\rangle}_{{\rm }}\left({l, b}\right) = A \exp \left({-\alpha \left|{b}\right|}\right)\hspace{2mm},
\end{eqnarray}
where $A$ and $\alpha$ are the fitting coefficients. We determined the coefficients ($A$, $\alpha$) to be (1.60, 0.03), (2.60, 0.05), and (0.84, 0.03) for $<$$B$$-$$V$$>$${}_{{\rm ref}}$,  $<$$V$$-$R$>$${}_{{\rm ref}}$, and  $<$$V$$-$$I$$>$${}_{{\rm ref}}$, respectively. \par
We then smoothed the color excess maps with a Gaussian filter (FWHM$=$4$'$) to reduce noise. The resulting resolution of the maps is typically $\sim$5$'$ in the reference field and $\sim$10$'$ in the most opaque region. An example of the obtained color excess maps is shown in Figure 2, where we show the spatial distribution of $E_{B-V}$. The distributions of $E_{V-R}$ and $E_{V-I}$ (not shown) are similar to that of $E_{B-V}$. L1251 is also traced in the all-sky $E_{B-V}$ map with 6$'$ resolution (Schlegel, Finkbeiner, \& Davis 1998), which is derived from far-infrared emission data ($COBE$/$DIRBE$, $IRAS$/$ISSA$). The cloud appears similar in both the all-sky map and our color excess maps, indicating that the optical color excess and the far-infrared thermal emission of dust trace the same region.
\par
Uncertainty in deriving $<$$E_{V-\lambda}$$>$ is given by following equation (e.g., Arce \& Goodman 1999):
\begin{eqnarray}
{\sigma }_{\left\langle{E\left({V -\ \lambda }\right)}\right\rangle} = { \left[ {{\sigma }_{{\rm cell}}^{2} + \sum\limits_{i = 1}^{N} {\left({{\frac{{\sigma }_{{\left({V -\ \lambda }\right)}_{i}}}{{N}^{}}}}\right)}^{2}} \right] }^{1 / 2}
\end{eqnarray}
where $\sigma_{{(V-\lambda)}_{i}}$ is the photometric error in $V$$-$$\lambda$ of the $i$th star in each cell. The value $\sigma_{{\rm cell}}$ is the standard deviation of the $N$-sampling distribution of the mean, where the original distribution is the $V$$-$$\lambda$ color of the stars in the reference field. It is equivalent to the standard deviation of the $<$$E_{V-\lambda}$$>$ distribution in the reference field. We estimated $\sigma_{{\rm cell}}$ to be 0.04, 0.03, and 0.05 mag for the color of $B$$-$$V$, $V$$-$$R$, and $V$$-$$I$, respectively. In the $E_{B-V}$, $E_{V-R}$, and $E_{V-I}$ maps, a typical $\sigma_{<{E}_{V-\lambda}>}$-value calculated using equation (6) is 0.05, 0.03, and 0.05 mag, respectively. \par
We derived the color excess diagram $E_{B-V}$ versus $E_{V-I}$ to see the global trend of reddening by dusts in L1251, which is shown in Figure 3. We investigated the relation separately in the star-forming head ($l$$>$114.$\hspace{-3.7pt}$${}^{\circ}$2) and in the non-star-forming tail ($l$$\le$114.$\hspace{-3.7pt}$${}^{\circ}$2). For comparison, we also show $E_{B-V}$ versus $E_{V-I}$ relations in the cases of $R_{V}$$=$3.1 and 6.5, adopting an empirical extinction law reported by Cardelli et al. (1989). \par
It is clear that there is a significant difference in the $E_{B-V}$ versus $E_{V-I}$ relations obtained in the two parts of the cloud. In the non-star-forming tail the data are distributed along the line of $R_{V}$$=$3.1, a typical value in the diffuse ISM, and the linear least-squares fit to the data actually yields a slope of $E_{V-I}$$/$$E_{B-V}$$=$1.21, which corresponds to $R_{V}$$=$3.2. On the other hand, the data in the star-forming head are distributed above the line of $R_{V}$$=$3.1 in the range of $E_{B-V}$$\ge$0.4 mag, and they gradually get close to the line of $R_{V}$$=$6.5 for larger $E_{B-V}$. It is evident that optical properties of dust grains significantly vary even in a single cloud. The different extinction laws are well separated into two groups only when we divide them into star-forming and non-star-forming parts. Since $R_{V}$ increases with increasing size of dust grains, larger $R_{V}$-value in the star-forming head is indicative of the evolution (i.e., growth in size) of dust grains. We note that $E_{V-I}$ increases in the same way (along with $R_{V}$$\sim$3.1) up to $E_{B-V}$$\sim$0.4 mag both in the tail and in the head part, which indicates dust optical properties remain normal at the diffuse boundary of the cloud.

\section{Derivation of $R_{V}$ Distribution and Evidence for Grain Growth}
In order to investigate the local variation of the dust properties in L1251, we derived the $R_{V}$ distribution using a relationship between $R_{V}$ and $E_{V-\lambda}$/$E_{B-V}$ that we produced from the $R_{V}$-dependent extinction law derived by Cardelli et al. (1989). They found that observed $A_{\lambda}$/$A_{V}$ versus $R_{V}^{-1}$ relations toward a number of line-of-sight stars are well fitted over the wavelength range from 0.125 $\mu$m to 3.5 $\mu$m as
\begin{eqnarray}
\left\langle{{A}_{\lambda } / {A}_{V}}\right\rangle = a(x) + b(x) {R}_{V}^{-1}\hspace{2mm},
\end{eqnarray}
where $x$ is ${\lambda}^{-1}$ $\mu$m${}^{-1}$. The wavelength-dependent coefficients $a(x)$ and $b(x)$ are shown in equations (2a-2b) and (3a-3b) of Cardelli et al. (1989). By using equation (7) we reproduced $R_{V}$ and $E_{V-\lambda}$/$E_{B-V}$ relations with the color excess ratios as $E_{V-\lambda}$/$E_{B-V}$=($A_{V}$-$A_{\lambda}$)$/$($A_{B}$-$A_{V}$). \par 
Color excess ratios of $E_{V-R}$/$E_{B-V}$ and $E_{V-I}$/$E_{B-V}$ are used to obtain the $R_{V}$-values at the cells that we used to derive the color excess maps. In each cell we numerically calculated the $R_{V}$-value, which minimizes the sum of residual variance; i.e., $\bigl[$($E_{V-R}$/$E_{B-V}$)${}_{{\rm obs}}$$-$($E_{V-R}$/ \\ $E_{B-V}$)${}_{{\rm model}}$$\}$${}^{2}$ $+$ $\{$($E_{V-I}$/$E_{B-V}$)${}_{{\rm obs}}$$-$($E_{V-I}$/$E_{B-V}$)${}_{{\rm model}}$$\bigr]$${}^{2}$. We carried out  the $R_{V}$ fitting only in the region with $E_{B-V}$$\ge$0.4 mag in order to avoid a large error in $R_{V}$ arising from too small $E_{B-V}$ values. The resulting $R_{V}$ map is shown in Figure 4. As seen in the figure, $R_{V}$ in the star-forming head is systematically higher than in the tail with the maximum value of $R_{V}$$\sim$6. The mean value of $R_{V}$ in the head and tail are $\sim$4.6 and $\sim$3.2, respectively. The feature in the $R_{V}$ distribution is consistent with what we found in the color excess diagrams derived in the previous section (Fig. 3).  \par
Uncertainties in the $R_{V}$ measurement, $\sigma$${}_{R_{V}}$, are estimated using Monte Carlo simulations. We first generated noise that follows a Gaussian distribution with the standard deviation equal to the 1 $\sigma$ error of $E_{V-\lambda}$$/$$E_{B-V}$ measurements, and then we performed the $R_{V}$ fitting to the noise-added $E_{V-\lambda}$$/$$E_{B-V}$ data. We repeated this calculation 1000 times in each cell, and regarded the standard deviation of the resulting $R_{V}$-values as $\sigma$${}_{R_{V}}$. The value of $\sigma$${}_{R_{V}}$ varies from $\sim$0.5 to $\sim$0.9 (0.69 on the average) and tends to be relatively higher in cells with small $E_{B-V}$. In addition to the random noise our $R_{V}$ map may suffer from a systematic error due to a number of unidentified pre-main-sequence (PMS) stars formed in the dense head. A certain fraction of PMS stars are known to have a blue color excess (e.g., Kenyon \& Hartmann 1995), which may result in an overestimate in $R_{V}$. However, this possible error should be negligible in our case, as we discuss in the Appendix.  \par
We further checked the validity of the color-excess-based $R_{V}$ measurements by using $A_{B}$ and $A_{V}$ derived with star count method. Though large uncertainties (typically $\sim$0.5 mag in our data) are included in $A_{B}$ and $A_{V}$ data, $R_{V}$ can be directly computed as $R_{V}$$=$$A_{V}$$/$$(A_{B}-A_{V})$ without using an empirical model. We made $A_{V}$ versus $A_{B}$ diagram for the same cells used in $R_{V}$ measurements and separated them into the star-forming head ($l$$>$114.$\hspace{-3.7pt}$${}^{\circ}$2) and non-star-forming tail ($l$$\le$114.$\hspace{-3.7pt}$${}^{\circ}$2). The slope of the $A_{V}$ versus $A_{B}$ relations yields mean $R_{V}$-value of data points as $<$$R_{V}$$>$$=$1$/$($\alpha$$-$1), where $\alpha$$=$${A}_{B}/{A}_{V}$. The linear least-squares fit to each dataset resulted in $\alpha$${}_{{\rm head}}$$=$1.23$\pm$0.09 and $\alpha$${}_{{\rm tail}}$$=$1.34$\pm$0.07, which correspond to $<$$R_{V}$$>$${}_{{\rm head}}$$\sim$4.3 and $<$$R_{V}$$>$${}_{{\rm tail}}$$\sim$3.0, respectively. These values are consistent with the results of color-excess-based $R_{V}$ measurements. \par
In Figure 4 it is particularly noteworthy that there is a good coincidence between the local peaks of the $R_{V}$ distribution and the locations of YSOs. There are three strong $R_{V}$ peaks in the head at ($l$,$b$)$=$(114.$\hspace{-3.7pt}$${}^{\circ}$60, 14.$\hspace{-3.7pt}$${}^{\circ}$47; $R_{V}$=6.1), (114.$\hspace{-3.7pt}$${}^{\circ}$50, 14.$\hspace{-3.7pt}$${}^{\circ}$67; $R_{V}$=5.8), and (114.$\hspace{-3.7pt}$${}^{\circ}$27, 14.$\hspace{-3.7pt}$${}^{\circ}$77; $R_{V}$=5.6). The first two peaks show a good agreement with the locations of the two outflow sources reported by Sato \& Fukui (1989) and Sato et al. (1994), though the third peak has no apparent corresponding source. In addition, the three $R_{V}$ peaks also coincide with the positions of molecular cloud cores found in C${}^{18}$O (named ``B", ``C", and ``E" by Sato et al. 1994; see their Table 1 and Fig. 2), indicating that the region with high $R_{V}$-values is rich in molecular gas.  \par
To investigate the dependence of $R_{V}$ on the dust column density, we derived a diagram of $A_{V}$ versus $R_{V}$. For a direct comparison we remeasured the $A_{V}$-values with the same cells used to produce the $R_{V}$ map. In Figure 5 we show the diagram measured both in the dense head and in the diffuse tail. It is clear that the $R_{V}$ increases along with $A_{V}$ in the head, while that in the tail remains constant regardless of $A_{V}$. \par
All of the above results obtained from the comparison of the $R_{V}$ map with the other data, i.e., the coincidence of the $R_{V}$ peaks and the outflow sources, as well as the C${}^{18}$O cores, and the difference in the $A_{V}$ versus $R_{V}$ relation between the head and the tail, strongly indicate that $R_{V}$ is higher in the dense star-forming head than those in the diffuse tail without any signs of star formation. Because a higher value of $R_{V}$ is expected for dusts with a larger size, the dusts in the head of L1251 may actually be larger than in the diffuse tail, indicating the growth of dust grains through coagulation and accretion of smaller dust particles as well as molecules in gas phase onto larger dust grains in the dense molecular cloud cores experiencing star formation. The $R_{V}$ distribution derived in this study may not be sufficient to probe the densest region of L1251 ($A_{V}$$>$4$-$5 mag) in detail because of a large optical depth at the observed wavelengths. However, a similar change of dust properties most likely representing a large dust size in molecular clouds was recently reported through submillimeter observations of the dust emission (e.g.,  MCLD123.5$+$24.9: Bernard et al. 1999; Taurus: Stepnik et al. 2003).  The idea of grain growth is also supported by their study based on the optically much thinner submillimeter observation, implying that dust grains with a larger size than in the diffuse ISM may be common in dense molecular clouds. \par
Our procedure to derive the $R_{V}$ distribution on the basis of the average color excess measurements is useful to investigate a large-scale variation of the dust optical properties. It may be applicable also to the near infrared data in the future, with which we should be able to measure the $R_{V}$ distribution in the densest region of L1251 to confirm the significant change of $R_{V}$ therein.
\section{Summary}
We have carried out optical observations toward L1251 at multiple bands, $B$, $V$, $R$, and $I$, using the 105 cm Schmidt telescope equipped with the 2kCCD camera at the Kiso Observatory in Japan. Extinction and color excess maps of the cloud were derived using a star count method and a NICE method. From the observed $E_{V-\lambda}$$/$$E_{B-V}$ ($\lambda$$=$$R$,$I$), we also derived the $R_{V}$ distribution using an empirical $R_{V}$ versus $A_{\lambda}$$/$$A_{V}$ relation (Cardelli et al. 1989). \par
(1) We investigated the color excess distribution of the cloud on the basis of the $\sim$30,600 stars detected in our observations. Color excess diagram, $E_{B-V}$ versus $E_{V-I}$, indicates a significant difference in $R_{V}$ between the star-forming dense head ($l$$>$114.$\hspace{-3.7pt}$${}^{\circ}$2) and the non-star-forming diffuse tail ($l$$\le$114.$\hspace{-3.7pt}$${}^{\circ}$2).  \par
(2) We derived the $R_{V}$  distribution over the cloud on the basis of the color excess measurements and an empirical relation between $R_{V}$  and $A_{\lambda}$$/$$A_{V}$ (Cardelli et al. 1989). From the resulting $R_{V}$ map we found that $R_{V}$ is systematically higher in the dense head ($R_{V}$$=$4$-$6) than in the diffuse tail ($R_{V}$$\sim$3.2), which is consistent with what we found in the $E_{B-V}$ versus $E_{V-I}$ diagram. We confirmed the validity of color-excess-based $R_{V}$ measurements by comparing them with independently measured $R_{V}$-values; i.e., $A_{V}$$/$($A_{B}$$-$$A_{V}$), which were directly computed using $A_{B}$ and $A_{V}$ data from star count analysis. \par

Local peaks of the $R_{V}$ distribution in the head of L1251 coincide with the positions of the young stars accompanied by molecular outflows and the molecular cloud cores identified in C${}^{18}$O. We also found that $R_{V}$ increases with increasing $A_{V}$ in the dense star-forming head, while $R_{V}$ is likely to be independent of $A_{V}$ in the diffuse tail. Since $R_{V}$ is closely related to the size of dusts, we suggest that these results are indicative of the growth of dust grains in the dense head due to coagulation and accretion of smaller dust particles and molecules onto the larger dust grains.

\acknowledgments

We are very grateful to M. Tamura, K. Tatematsu, and T. Mizuno for their helpful comments and suggestions. Thanks are due to the staff at Kiso Observatory including K. Nakada, T. Miyata, T. Soyano, K. Tarusawa, Y. Tanaka, H. Mito, T. Aoki, N. Ito, and S. Nishiura for their kind support during the observations. This work was financially supported by the Grant-in-Aid for Scientific Research by the Japanese Ministry of Education, Science, Sports and Culture (Nos. 13640233 and 14022214).

\appendix

\section{A Possible Systematic Uncertainty in $R_{V}$}
Here we consider the degree of uncertainties caused by the presence of unknown faint PMS stars in the star-forming head of L1251, which escaped detection in the previous H${}_{\alpha}$ survey (Kun \& Prusti 1993). It is well known that a certain fraction of PMS stars show the optical blue excess, which is prominent in $U$$-$$B$ (e.g., Kenyon \& Hartmann 1995) and possibly in $B$$-$$V$. Their blue intrinsic color should result in an underestimate of $E_{B-V}$, for which we may overestimate $R_{V}$. \par
In order to examine the fraction of blue excess PMS stars at $B$$-$$V$, we made a $B$$-$$V$ versus $V$$-$$I$ diagram of 96 PMS stars in the Taurus-Auriga region (see Table A1 of Kenyon \& Hartmann 1995). We found 23 ($\sim$25$\%$) sources show a blue excess of greater than 0.2 mag compared with the main-sequence stars, while the rest of the sources have a similar color to the main-sequence stars with some reddening.  \par
If we assume that the ratio of the blue excess PMS stars measured in the Taurus-Auriga region is also valid in L1251, only two to three stars (25$\%$ of $N$$=$10) in each counting cell are expected to be blue excess PMS stars even in an extreme case that all of the stars detected in our observations are PMS stars. However, at the maximum estimate of the blue excess PMS stars (about two to three stars per cell), there might be an influence on the value of $E_{B-V}$ calculated as the average color of 10 stars (eq. [4]). We therefore recalculated $E_{B-V}$ taking the median color of 10 stars instead of the average, because it should be much less affected by a few blue excess PMS stars. We compared the two values of $E_{B-V}$, the median and the average colors of 10 stars, and found that they are quite consistent with each other even in the densest head of L1251 with a typical difference of only $\sim$0.02 mag. This means that the possible error in $R_{V}$ due to the blue excess PMS stars is negligible, and the high $R_{V}$-value derived in the star-forming dense head is most likely to be real.

\clearpage

\begin{figure}
\plotone{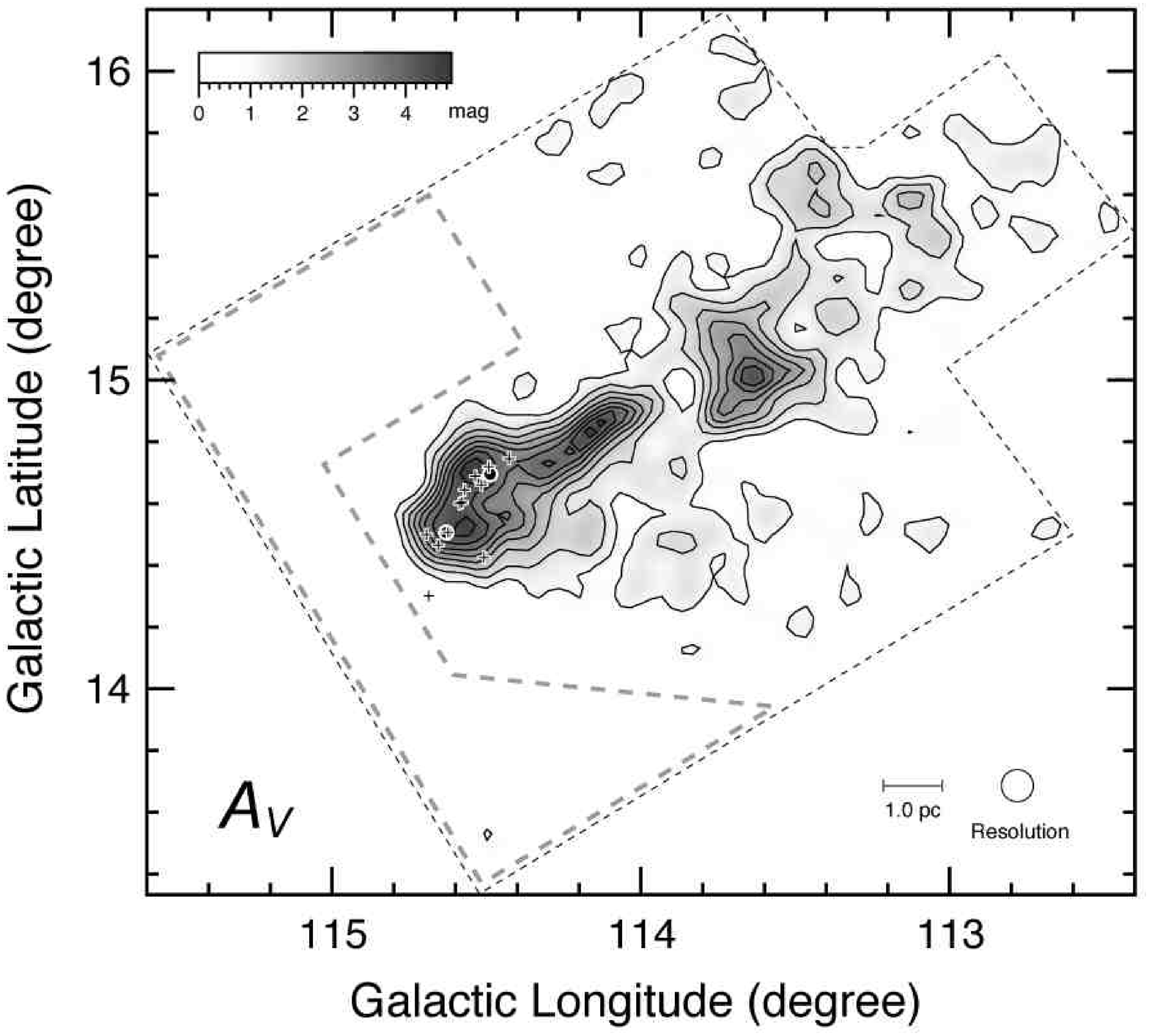}
\caption{$A_{V}$ distribution in L1251. The angular resolution is 6$'$. Contours start from 1 mag with a step of 0.5 mag. Filled circles with white outlines represent the positions of the $IRAS$ sources associated with molecular outflows (Sato et al. 1994). Plus signs represent the positions of the H${}_{\alpha}$ emission-line stars found by Kun \& Prusti (1993). The observed area is delineated by the broken line. The area enclosed by the gray broken line is the reference field expected to be free from extinction. Typical uncertainties in $A_{V}$ are $\sim$0.35 mag and $\sim$0.7 mag, which are found in the reference field ($A_{V}$$\sim$0 mag) and in the most opaque region ($A_{V}$$\sim$5 mag), respectively. \label{fig1}}
\end{figure}

\clearpage 

\begin{figure}
\plotone{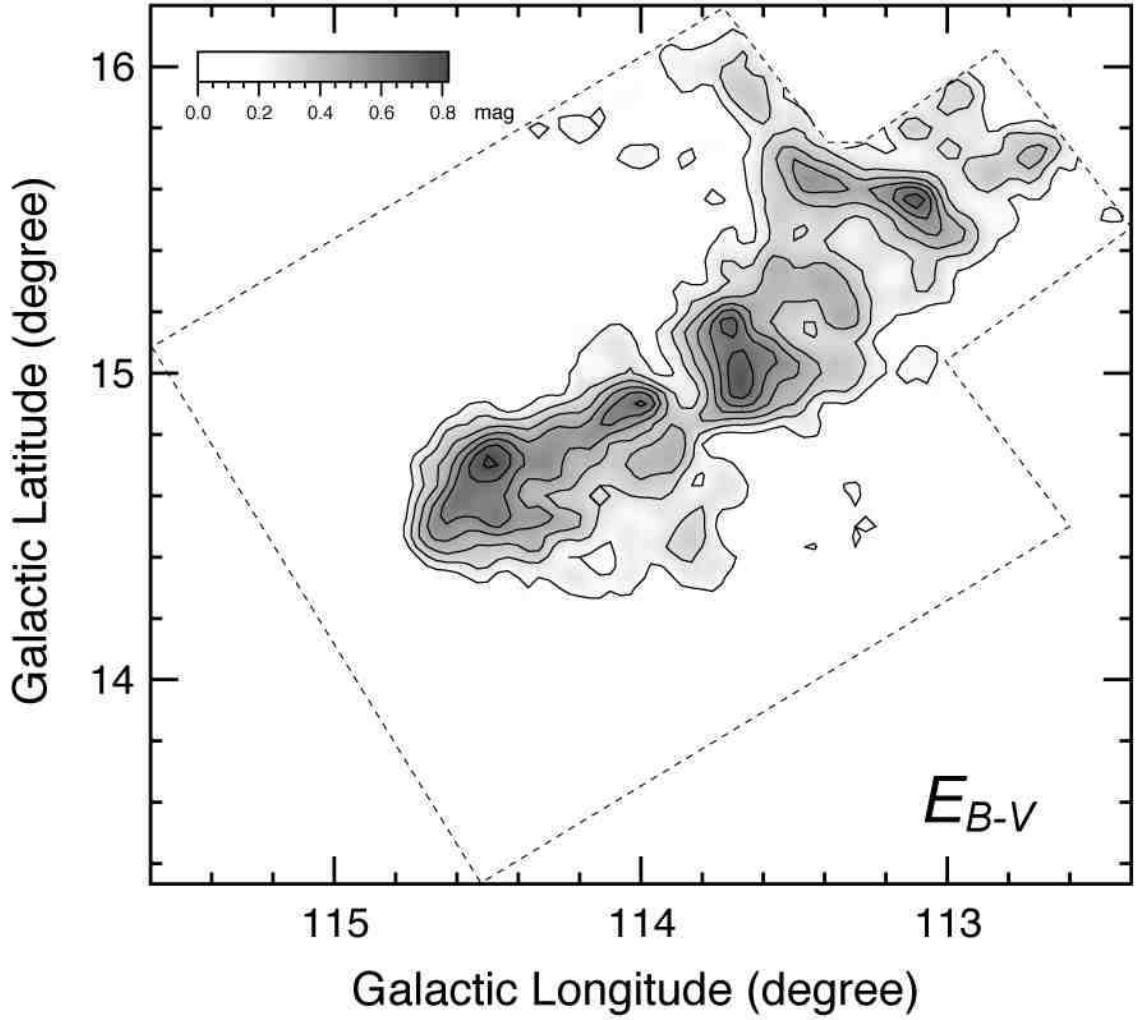}
\caption{$E_{B-V}$ distribution in L1251. Contours start from 0.2 mag with a step of 0.1 mag. The uncertainty in $E_{B-V}$ is $\sim$0.05 mag. The highest and lowest angular resolutions achieved in the map are $\sim$5$'$ in the reference field and $\sim$10$'$ in the most opaque region, respectively. \label{fig2}}
\end{figure}

\clearpage 

\begin{figure}
\plotone{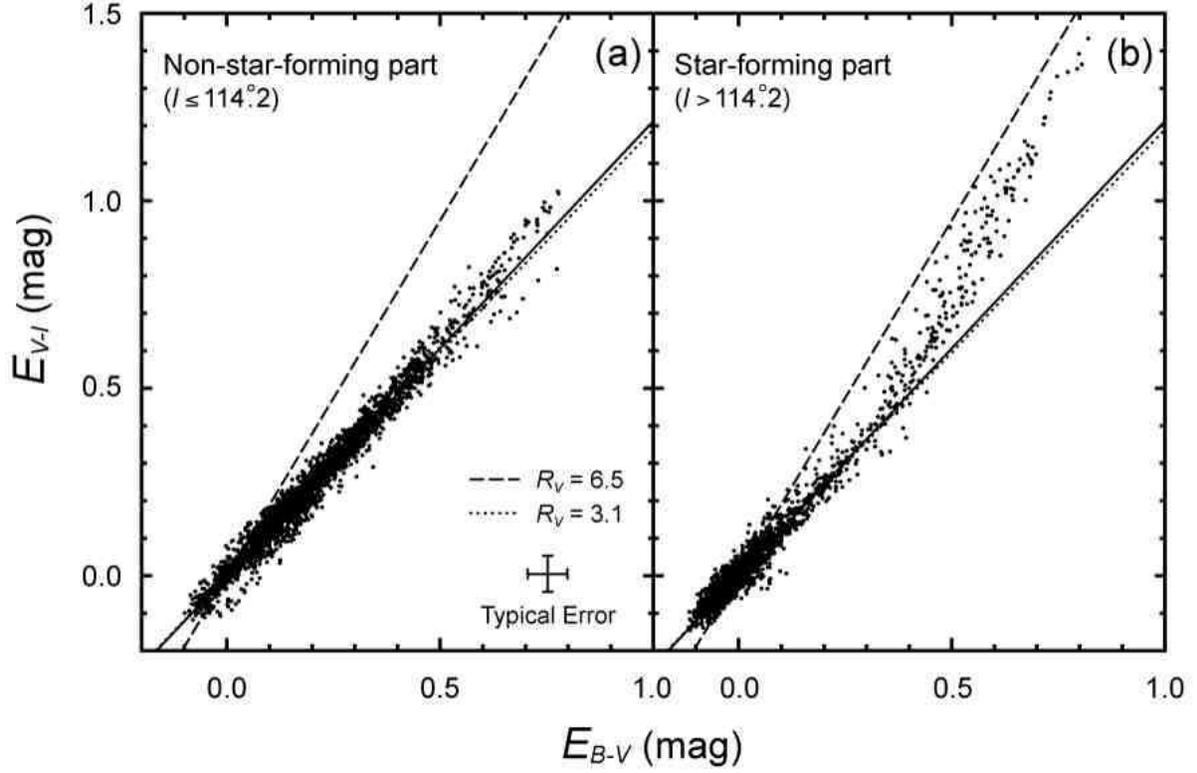}
\caption{(a) $E_{B-V}$ vs. $E_{V-I}$ relation for the non-star-forming tail of L1251 ($l$$\le$114.$\hspace{-3.7pt}$${}^{\circ}$2). The solid line denotes the linear least-squares fit ($E_{V-I}$$/$$E_{B-V}$$=$1.21 corresponding to $R_{V}$$=$3.2) to all of the data points in the tail. The dotted and broken lines are for the cases of $R_{V}$$=$3.1 and 6.5, respectively. The typical error in $E_{B-V}$ and $E_{V-I}$ is $\sim$0.05 mag. (b) Same as (a) but for the star-forming head of L1251 ($l$$>$114.$\hspace{-3.7pt}$${}^{\circ}$2). \label{fig3}}
\end{figure}

\clearpage 

\begin{figure}
\plotone{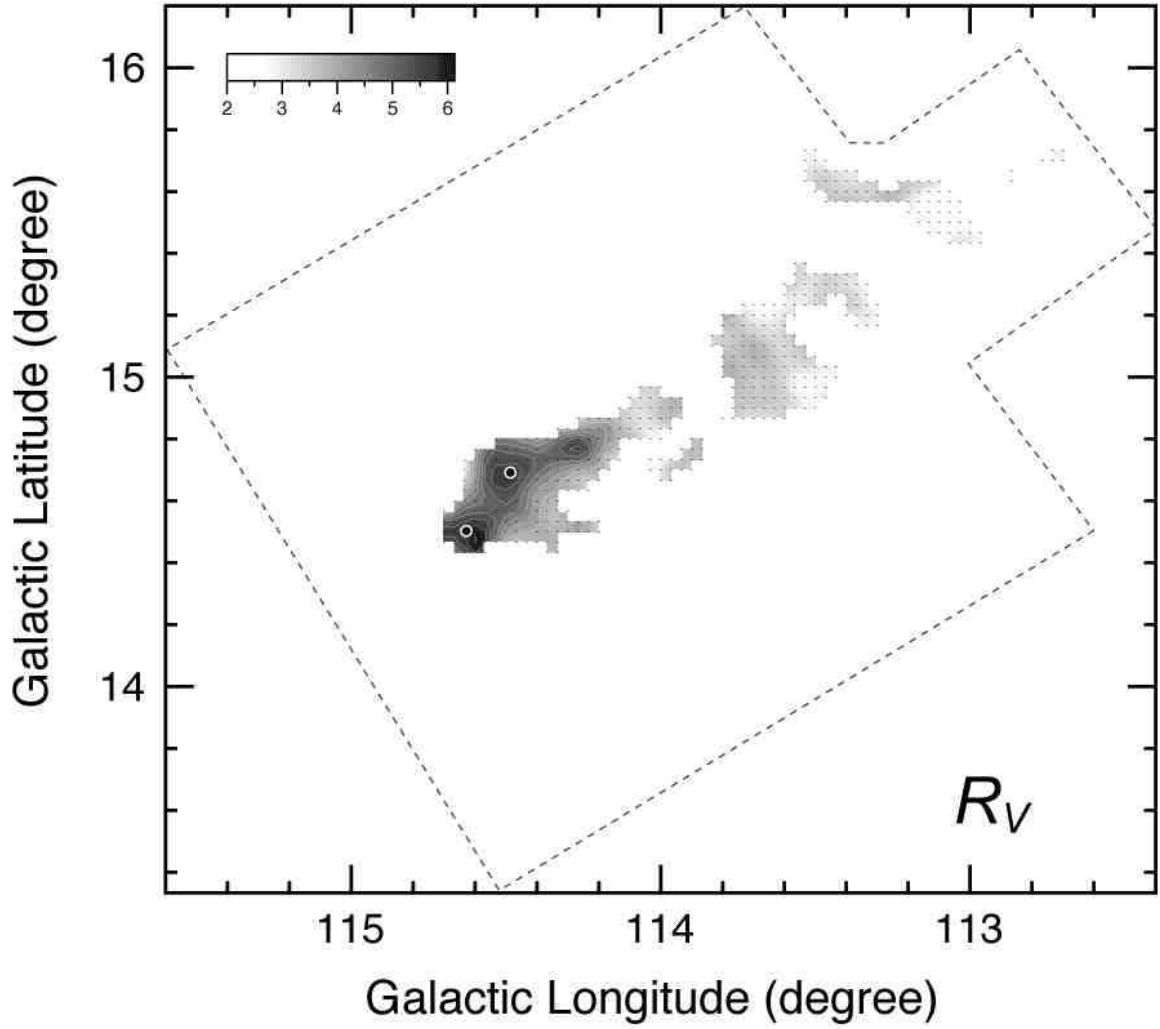}
\caption{$R_{V}$ distribution in L1251. The small dots represent the positions where we measured the $R_{V}$-values. Filled circles with white outlines denote the $IRAS$ sources associated with molecular outflows (Sato \& Fukui 1989; Sato et al. 1994). Contours start from $R_{V}$=4 with a step of 0.3. \label{fig4}}
\end{figure}

\clearpage 

\begin{figure}
\plotone{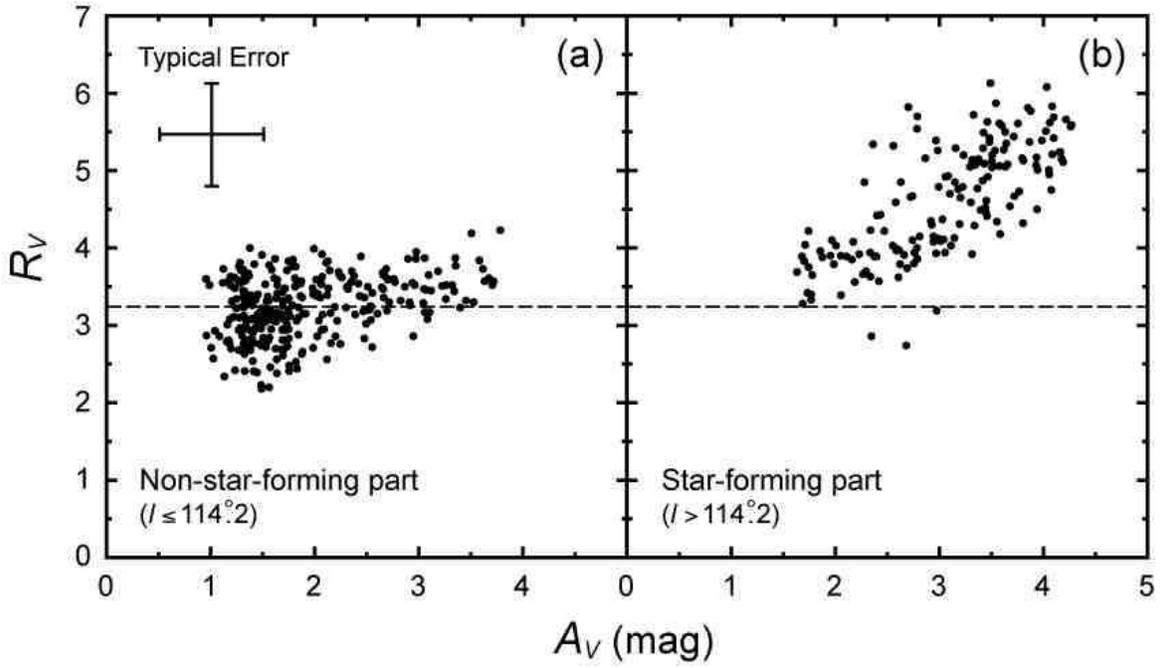}
\caption{(a) $A_{V}$ vs. $R_{V}$ relation for the non-star-forming tail of L1251 ($l$$\le$114.$\hspace{-3.7pt}$${}^{\circ}$2). The broken line denotes the mean $R_{V}$-value 3.2. A typical error in $A_{V}$ and $R_{V}$ is 0.5 mag and 0.69, respectively. (b) Same as (a) but for the star-forming head of L1251 ($l$$>$114.$\hspace{-3.7pt}$${}^{\circ}$2). \label{fig5}}
\end{figure}

\end{document}